\documentclass[12pt,twocolumn]{article}
\usepackage[a4paper,left=20mm,right=20mm,top=25mm,bottom=25mm,includeheadfoot]{geometry}

\usepackage{mathpazo}
\usepackage{graphicx}
\usepackage{amsmath,textcomp}
\usepackage{makeidx}
\usepackage{graphicx}% Include figure files
\usepackage{dcolumn}% Align table columns on decimal point
\usepackage{bm}% bold math
\usepackage{fancyhdr}
\usepackage{natbib}
\usepackage{color}

\makeindex

\usepackage{sectsty}
	\allsectionsfont{\sffamily\raggedright}
	\sectionfont{\sffamily\large\raggedright}
	\subsectionfont{\sffamily\normalsize\raggedright}
\usepackage{ftnright}

\usepackage{widetext}
\usepackage{flushend}
\usepackage{cuted}
\usepackage{color}
\usepackage{graphicx}
\usepackage{hyperref}
\usepackage{pstricks,amssymb}

\usepackage{fancyhdr}
\begin{document}
%Do not change the \vspace command, which is needed to suppress extra space above the title.
\title{\vspace{-2em}\bfseries\sffamily Intrinsic nonlinearity of a PN-junction diode and higher order harmonic generation}
\author{\normalsize Samridhi Gambhir, Arvind, Mandip Singh\\[2ex]
Department of Physical Sciences,\\Indian Institute of Science Education and Research Mohali, India.\\
{\tt mandip@iisermohali.ac.in}
}

\date{\itshape}

\maketitle

\thispagestyle{fancy}

%Do not change the \sffamily command, it is needed to ensure that the abstract appears in a different font.
\begin{abstract}
{\sffamily
Voltage current characteristics of a $PN$-junction diode are intrinsically nonlinear in nature. It is shown in this paper that a mathematical form of nonlinearity of a $PN$-junction diode resembles the nonlinear response of electric polarization of a dielectric medium to the electric field. Nonlinearity of a $PN$-junction can be expressed in a series of successively increasing orders of the nonlinearity. For a $PN$-junction diode, higher order nonlinear terms become significant as a voltage across the diode is increased. In this paper, a gradual emergence of a nonlinear regime with the amplitude of a sinusoidal voltage is presented. Higher order harmonics are produced by utilizing the nonlinearity of a single $PN$-junction diode. An experimental realization of a frequency comb with the highest frequency up to the twentieth harmonics is also presented. In addition, in the same circuit by making the nonlinearity significant up to the second order, an experiment on generation of the sum and difference of frequencies is realized.}
\hrule
%Do not change the \hrule command, it is needed to separate the abstract form the main text.
\end{abstract}

\section{Introduction}
Nonlinearity is present in most of the natural processes. A linear behaviour of a system is an approximation of a generalised nonlinear response [~\cite{nonlinear}]. Whether it is a simple pendulum or interaction of light with matter, a large amplitude of oscillation leads to a breakdown of a linear approximation and brings out the nonlinear response of a system.  Nonlinear physics is a well established field of research. Some of the prominent concepts such as the emergence of chaos [~\cite{chaos}] and in nonlinear optics where light fields interact with each other [~\cite{shen,boyd, solitons}] have been of great interest. In this paper, an experiment based on a $PN$-junction diode is introduced which utilizes the intrinsic nonlinearity of a $PN$-junction diode to experimentally realize the effects analogous to those observed in classical nonlinear optics. The voltage current characteristics of a $PN$-junction diode can be expressed in the form of an infinite series of linear and nonlinear terms. With an increase in the amplitude of a sinusoidal voltage applied across a $PN$-junction diode the higher order nonlinearities become important. As the amplitude of a sinusoidal voltage is increased, new components of frequency appears in current passing through the $PN$-junction diode. A gradual generation of new frequencies and generation of a frequency comb [~\cite{hanschreview}] with the highest frequency up to the twentieth harmonics are clearly demonstrated in the experiment.  Further, by using a source of two different frequency sinusoidal voltage waveforms (a sinusoidal voltage has a wave like form in the time domain), the sum and  difference frequency generation have been demonstrated experimentally.

\section{Nonlinear dielectric medium: Introduction}
In a linear dielectric medium in presence of an electric field, the displacement of bound charges from their equilibrium position remains linearly proportional to the applied force (Hooke's law). As a consequence, the induced electric polarization ($P$) remains linearly proportional to the net electric field ($E$) such that $P=\displaystyle \epsilon_{0}\chi_{e} E$, where $\epsilon_{0}$ is the vacuum permittivity and $\chi_{e}$ is the electric susceptibility, which is a measure of the ability of a medium to get polarized.  Therefore, when light propagates in a `linear' dielectric medium, the frequency of the propagating light wave and the refractive index of the medium remain unaltered. As a consequence, different electromagnetic fields do not interact with each other.  However, a linear response is an approximation and as the intensity of light (electric field) is increased, the displacement of bound charges increases.  Hooke's law does not remain valid for a large displacement of bound charges and hence the induced polarization does not remain linearly proportional to the electric field.  The dependence of the induced electric polarization on the electric field becomes nonlinear which can be expressed as [\cite{shen,photonics,boyd}]

\begin{equation} \label{eq:1}
P=\epsilon_{0}(\chi^{(1)}_{e}E+\chi^{(2)}_{e}E^{2}+\chi^{(3)}_{e} E^{3} +\cdot\cdot\cdot)
\end{equation}

Where $\displaystyle \chi^{(1)}_{e}$,  $\chi^{(2)}_{e}$ and $\displaystyle \chi^{(3)}_{e}$ are the first, second and third order
electric susceptibilities respectively, the order of nonlinearity is denoted by a superscript. Equation.~\ref{eq:1} can be  rewritten as  the sum of a linear and a nonlinear term such that $P=\displaystyle \epsilon_{0}\chi^{(1)}_{e}E + P_{NL}$, where $P_{NL}$ is the nonlinear component of  the electric polarization. At a low light intensity, only the first term is significant and all the nonlinear terms are negligible. For a large intensity of light, the $P_{NL}$  term becomes pronounced. The relative magnitude of  the nonlinear susceptibilities in  Equation~\ref{eq:1} depends on the crystal symmetry [\cite{ajp-scale}]. For non centro-symmetric crystals such as Lithium Niobate (LiNbO$_{3}$), Potassium Titanyl Phosphate (KTP) and Beta Barium Borate (BBO) the second order nonlinearity $\displaystyle \chi^{(2)}_{e}$ is nonzero. A nonlinear medium having only the second order nonzero nonlinearity is known as a second order nonlinear medium [\cite{shen, photonics, boyd}]. One of the important processes resulting from the second order nonlinearity ($\displaystyle \chi^{(2)}_{e}$-process) is the second harmonic generation [~\cite{second}], where frequency of light is doubled. According to the quantum description of second harmonic generation, two photons of frequency $\omega$ interact with each other via the second order interaction and merge together as a single photon of frequency $2\omega$. In a reverse process which is known as frequency down conversion, a photon can split into two photons of lower frequencies. Other examples of second order nonlinear processes are the sum and  difference frequency generation [~\cite{sum-1,shenreview, shen, photonics, boyd, frankenreview, hanschreview}].  For a medium such as the silica optical fiber, $\chi^{(3)}_{e}$ is significant which results  in third order nonlinear effects [~\cite{thyagarajan, agarwal}], where the refractive index becomes intensity dependent - a phenomenon known as the optical Kerr effect. Optical Kerr effect can be utilised to modulate the phase  of a wave by modulating the intensity of another wave. A few important examples of third order nonlinear processes [~\cite{photonics, agarwal}] are the third harmonic generation [~\cite{third}], self phase modulation, cross phase modulation, four wave mixing, supercontinuum generation and optical phase conjugation. These nontrivial effects have revolutionized the field of nonlinear optics over the decades and have played a significant role to produce quantum entangled photons and in experiments on quantum information processing.

%%%%%%%%%%%%%%%%%%%%%%%%%%%%%%%%%%%%%%%%%%%%%%%%%%%%%%%%%
\section{Nonlinearity of a PN-junction diode}
\label{pn-junction}
An ideal diode conducts current in the forward bias and completely blocks the flow of current in the reverse bias. A $PN$-junction is formed, if a $P$-type semiconductor is joined to a $N$-type semiconductor. Immediately after joining them, electrons which are majority carriers of $N$-type region start diffusing into $P$-type region due to their concentration gradient across the contact of two semiconductors. Diffused electrons recombine with holes which are majority carriers of $P$-type region. Flow of charge carries through the contact due to concentration gradient generates a diffusion current.  However, this process is gradually stopped by an opposing force. As electrons diffuse into the $P$-type region, a net positive charge develops near the contact in $N$-type region and a net negative charge is formed near the contact in $P$-type region. This charge distribution produces an electric field in the $PN$-junction pointing from $N$-type region towards $P$-type region. Electric field generates an electric potential barrier at the junction and an opposing force for the flow due to diffusion of majority carriers. However, minority carriers which are produced by thermal agitation, are accelerated by the junction electric field. This process produces current named as the drift current. In equilibrium condition, if a $PN$-junction is unbiased, the diffusion current and drift current flow in opposite direction. Therefore, a net flow of current is zero through a $PN$-junction \emph{i.e.} majority carriers climb the potential barrier and minority carriers descends the potential barrier.  Under the forward bias, $P$-type region is kept at a higher potential as compared to $N$-type region by connecting to a battery. In the forward bias, the net electric field and height of potential barrier across the junction is decreased. Therefore, probability of carriers to climb the potential barrier increases in the forward bias. This process increases electron concentration in $P$-type region at the junction edge and hole concentration in $N$-type region at the junction edge. Such an excess carrier concentration increases exponentially with an applied bias voltage across the junction. Therefore, in the forward bias, the diffusion current increases exponentially with applied bias voltage while the drift current remains negligibly small. In the reverse bias, $P$-type region is kept at a lower potential as compared to $N$-type region. Therefore, potential barrier height increases and probability to cross the potential barrier decreases. However, the flow of drift current saturates with the increase in the reverse bias voltage due to an extremely low carrier concentration of thermally generated minority carriers. The flow of net current through a $PN$-junction is a nonlinear function of the bias voltage. However, in case of a nonlinear dielectric medium, it is the displacement of charge from the equilibrium position which depends nonlinearly on the applied force. Nonlinear dielectric medium and a $PN$-junction diode are analogous to each other in the context of a mathematical form of their nonlinearity, which is the central idea of this paper.   

In a  realistic $PN$-junction diode, for a bias voltage $V$ across a junction the net current $I$ passing through the junction is [~\cite{electronics}]
\begin{equation}
\label{eq:2}
I=I_{0}(e^{V/\eta V_{th}}-1)
\end{equation}

Where $I_{0}$ is the reverse saturation current, $\eta$ is the ideality factor which is constant (value between 1 to 2) and $V_{th}=\frac{k_{B} T}{q}$ is the thermal voltage, $k_{B}$ is  the Boltzmann constant, $T$ is the absolute temperature and $q$ is the charge on electron. A typical numerical value of the
thermal voltage is $25.85~mV$ at $300~K$. According to Equation.~\ref{eq:2}, a $PN$-junction diode current increases exponentially when a forward bias voltage $V$ is increased, while under a reverse bias the
current saturates to $I_{0}$.
\begin{center}
	\begin{figure}[ht]
		\includegraphics[scale=0.5]{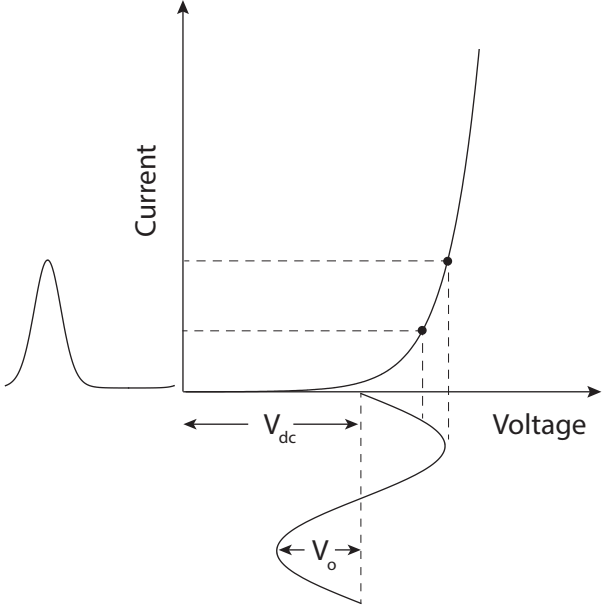}
		\caption{\label{fig:fig1} A schematic showing an
			exponential dependacnce of current on the forward bias
			voltage across a $PN$-junction diode. A variation of diode current for a
			sinusoidal bias voltage of amplitude $V_{o}$ with a dc offset
			$V_{dc}$ is also shown.}
	\end{figure}
\end{center}
Taylor series expansion of
Equation.~\ref{eq:2} can be written as
\begin{multline}
\label{eq:3}
I= I_{0}\bigg(\frac{V}{\eta
	V_{th}}+\frac{1}{2!}\frac{V^{2}}{\eta^{2} V^{2}_{th}}+
\frac{1}{3!}\frac{V^{3}}{\eta^{3}
	V^{3}_{th}}
+\\\frac{1}{4!}\frac{V^{4}}{\eta^{4}V^{4}_{th}}+
\frac{1}{5!}\frac{V^{5}}{\eta^{5} V^{5}_{th}}+\ldots \bigg)
\end{multline}
Which can be expressed as
\begin{multline}
\label{eq:4}
I =\chi^{(1)}_{v} V+\chi^{(2)}_{v} V^{2} +\chi^{(3)}_{v}
V^{3}+\\\chi^{(4)}_{v} V^{4}+\chi^{(5)}_{v} V^{5}+ \ldots
\end{multline}
Where $\displaystyle \chi^{(1)}_{v}=I_{0}/\eta V_{th}$,
$\displaystyle \chi_{v}^{(2)}=I_{0}/2!\eta^{2} V^{2}_{th}$,
$\displaystyle \chi^{(3)}_{v}=I_{0}/3!\eta^{3} V^{3}_{th}$,
and so on. Where a superscript denotes the order of
nonlinearity. The first term of Equation.~\ref{eq:4} is
linear and all the remaining terms represent a nonlinear
dependence of current on voltage.
Equation.~\ref{eq:4} is analogous to Equation.~\ref{eq:1} where the voltage is
analogous to the electric field and the current is analogous
to the electric polarization.

\begin{center}
	\begin{figure}
		\begin{center}
			\includegraphics[scale=0.47]{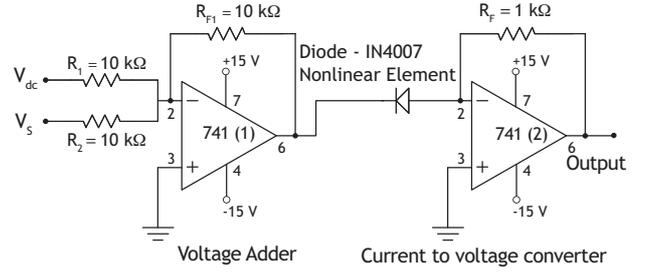}
			\caption{\label{fig:fig2} A diagram of a circuit designed to study the nonlinearity of a $PN$-junction diode. Two 741 general purpose operational amplifiers with a $PN$-Junction diode IN4007 are used in the circuit. Output of an inverting voltage adder is applied across a $PN$-junction diode. The output voltage of the circuit is proportional to current passing through the $PN$-junction diode.}
		\end{center}
	\end{figure}
\end{center}

\begin{center}
	\begin{figure}[ht]
		\begin{center}
			\includegraphics[scale=0.4]{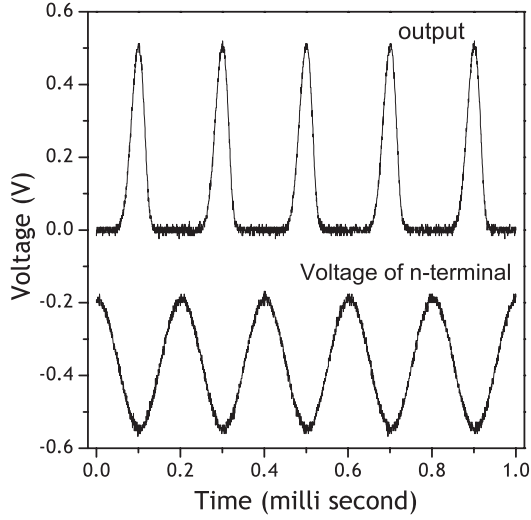}
			\caption{\label{fig:fig3} A plot showing an applied sinusoidal voltage across a
				$PN$-junction diode (lower plot).  Output voltage $V_{t}$ (upper plot) is
				linearly proportional to current passing through the
				diode. It is clearly shown that a sinusoidal input voltage gives rise to
				a  nonsinusoidal output voltage due to the intrinsic nonlinearity of the diode.}
		\end{center}
	\end{figure}
\end{center}

The voltage current characteristics of a $PN$-junction diode are qualitatively presented in Figure.~\ref{fig:fig1} where an exponential rise of current with the bias voltage is shown. A sinusoidal voltage of amplitude
$V_{o}$ with a constant dc-offset voltage $V_{dc}$ is applied across the diode. Because of the nonlinear characteristics of the $PN$-junction diode the current is  nonsinusoidal. For a positive dc-offset voltage such that $V_{dc}>V_{o}$, the diode is always forward biased. However, for $V_{dc}=0$ the diode is reverse biased during the negative half cycle and consequently a half wave rectification occurs. A circuit designed to study the nonlinearity of a $PN$-junction diode is shown in Figure.~\ref{fig:fig2}. This circuit produces a voltage output which is linearly proportional to current passing through the diode. Input voltage of the circuit is applied parallel to the diode. Circuit consists of two 741 general purpose operational amplifiers and a $PN$-junction diode IN4007 which is a nonlinear element in the circuit. The first operational amplifier is configured in an inverting voltage adder configuration whose output voltage is a summation of the input voltages with reversed polarity ($-(V_{s}+V_{dc})$). The output of the voltage adder is connected to the N-type terminal of a $PN$-junction diode (IN4007) while the P-type terminal is connected to the inverting terminal of a second operational amplifier. A typical numerical value of reverse saturation current $I_{o}$ of the $PN$-junction diode-IN4007 is $5.0~\mu A$. Since an inverting terminal of the second operational amplifier is biased at the virtual ground potential, therefore, the $PN$-junction diode becomes forward biased if the output of the voltage adder is negative. The second operational amplifier is operating in an inverting current-to-voltage converter configuration and its output voltage is $V_{t}=R_{F}I$, where $I$ is the current flowing through the $PN$-junction diode. Therefore, the output voltage of the current-to-voltage converter is $V_{t}=R_{F}I_{0}(e^{(V_{s}+V_{dc})/\eta V_{th}}-1)$. Final output voltage $V_{t}$ of the circuit and voltage across  the $PN$-Junction diode (potential at the N-type terminal where the P-type terminal is connected to a virtual ground) is shown in Figure.~\ref{fig:fig3}, where $V_{dc}=~$360mV and $V_{o}=~$180mV.

Nonlinear response of the circuit is clearly shown in the upper plot of Figure.~\ref{fig:fig3} where the current passing through the diode (converted to voltage by an operational amplifier) is non sinusoidal for a sinusoidal input voltage. The output voltage waveform is non-sinusoidal therefore, it shows the presence of more than one frequency components.

%%%%%%%%%%%%%%%%%%%%%%%%%%%%%%%%
\section{Phase Matching}
In nonlinear wave mixing processes, conditions of energy and momentum conservation should be fulfilled. The first condition implies that the total energy of interacting fields must be same before and after the nonlinear interaction. If all the waves at different spatial points at an instant of time in an extended medium are in phase with each other then a constructive interference occurs. This results in a high efficiency of harmonic generation. Any deviation from an exact phase matching results in a decrement of efficiency of harmonic generation. A phase matching condition implies conservation of momentum in an extended nonlinear medium. In case of a $PN$-junction diode, the nonlinearity is not extended however, its source of origin is localized at the junction. For a typical frequency of oscillating voltage waveform the wavelength of oscillating voltage waveform in circuit is much larger than the extension of the circuit. A nonlinear $PN$-junction is equivalent to a single nonlinear dipole and in this case the condition of phase matching implies resonance of driving field with the oscillator modes. In case of an extended nonlinear medium the nonlinearity is a collective process and coefficients of the orders of nonlinearity are governed by symmetry of a nonlinear crystal however, in case of a single dipole the notion of symmetry disappears. It is shown in this paper that all nonlinear terms appear for a $PN$-junction diode in a nonlinear regime. 

\section{Emergence of Nonlinear Regime}
\label{demo}
We now move to experiment to demonstrate a generation of new frequencies from 
the nonlinear response of the $PN$-junction diode. The output voltage $V_{t}$ of the circuit shown in
Figure.~\ref{fig:fig2} can be written as
\begin{multline}
\label{eq:5}
V_{t}=R_{F}(\chi^{(1)}_{v} V+\chi^{(2)}_{v} V^{2}
+\chi^{(3)}_{v} V^{3} +\chi^{(4)}_{v} V^{4} +\\
\chi^{(5)}_{v}
V^{5}+ \ldots)
\end{multline}
Where a voltage applied at input of circuit is $V=
V_{dc}+V_{s}$ and $V_{s}=V_{o} \cos(\omega_{o}t)$ is a
sinusoidal voltage corresponding to a fundamental
harmonic of angular frequency $\omega_{o}$. For a low
input voltage $V_{o}+V_{dc}$ the linear term is the only
significant term, therefore, the output voltage $V_{t}$ is
$R_{F}\chi^{(1)}_{v}\left(V_{dc} +
V_{o}\cos(\omega_{o}t)\right)$, which has the same frequency
spectrum as the spectrum of the input waveform,  namely the
presence of a single frequency $\omega_0$.
If the amplitude of the input sinusoidal voltage $V_{o}$ is
increased such that only the second order term of
Equation.~\ref{eq:5} becomes significant and all the
remaining nonlinear terms are negligible then the total
output voltage due to the first two terms of
Equation.~\ref{eq:5} can be written as
\begin{multline}
\label{eq:6}
V_{t}=R_{F}\bigg(\chi^{(1)}_{v} V_{dc} + \chi^{(2)}_{v}
\big(V^{2}_{dc}
+\frac{V^{2}_{o}}{2}\big)+\\(\chi^{(1)}_{v}+2\chi^{(2)}_{v}
V_{dc}) V_{o} \cos (\omega_{o}t) +\\\chi^{(2)}_{v}
\frac{V^{2}_{o}}{2} \cos(2\omega_{o}t)\bigg)
\end{multline}

The voltage waveform $V_{t}$ which is proportional to the
current passing through the $PN$-junction diode contains
a new frequency component ($2\omega_{o}$) of frequency
twice the frequency ($\omega_{o}$) of the input sinusoidal
voltage $V_{s}$. The generation of new frequency component corresponds to the second harmonic
generation.  A further increase of amplitude $V_{o}$ 
reinforces the third order nonlinear term to be significant in addition to the second order term.
The contribution to the output voltage $V_{t}$ due to the
third term of Equation.~\ref{eq:5} is written as
\begin{multline}
\label{eq:7}
R_{F}\chi^{(3)}_{v}\bigg(V^{3}_{dc} + \frac{3}{2} V_{dc}
V^{2}_{o}+ 3\big(V^{2}_{dc}
V_{o}+\frac{V^{3}_{o}}{4}\big)\cos(\omega_{o}t)\\+
\frac{3}{2} V_{dc}V^{2}_{o} \cos(2\omega_{o}t) +
\frac{V^{3}_{o}}{4} \cos(3\omega_{o}t)\bigg)
\end{multline}
which has frequency components at zero, $\omega_{o}$,
$2\omega_{o}$ and $3\omega_{o}$.  The harmonics of frequency
$3\omega_{o}$ corresponds to a third harmonic generation. The total output voltage $V_{t}$  is a
summation of contribution from the first three terms of
Equation.~\ref{eq:5}.  Similarly, the contribution from the
fourth order nonlinear term of Equation.~\ref{eq:5} to the
output voltage $V_{t}$ is written as
\begin{multline}
\label{eq:8}
R_{F} \chi^{(4)}_{v} \bigg(V^{4}_{dc} + 3 V^{2}_{dc}
V^{2}_{o} +  \frac{3 V^{4}_{o}}{8} + (4 V^{3}_{dc} V_{o}+ \\3
V_{dc} V^{3}_{o})\cos (\omega_{o}t)+  \big(3
V^{2}_{dc} V^{2}_{o}+  \frac{V^{4}_{o}}{2}\big)
\cos(2\omega_{o}t) \\+  V_{dc} V^{3}_{o}\cos {(3\omega_{o}
	t)}+ \frac{V^{4}_{o}}{8} \cos(4\omega_{o}t)\bigg)
\end{multline}
which has frequency components at zero, $\omega_{o}$,
$2\omega_{o}$, $3\omega_{o}$ and $4\omega_{o}$. The
frequency component at $4\omega_{o}$ signifies the fourth
harmonic generation. The total output voltage $V_{t}$
is again an addition of contributions from all the significant
terms  in  Equation.~\ref{eq:5}.
\begin{figure*}
		\includegraphics[scale=0.42]{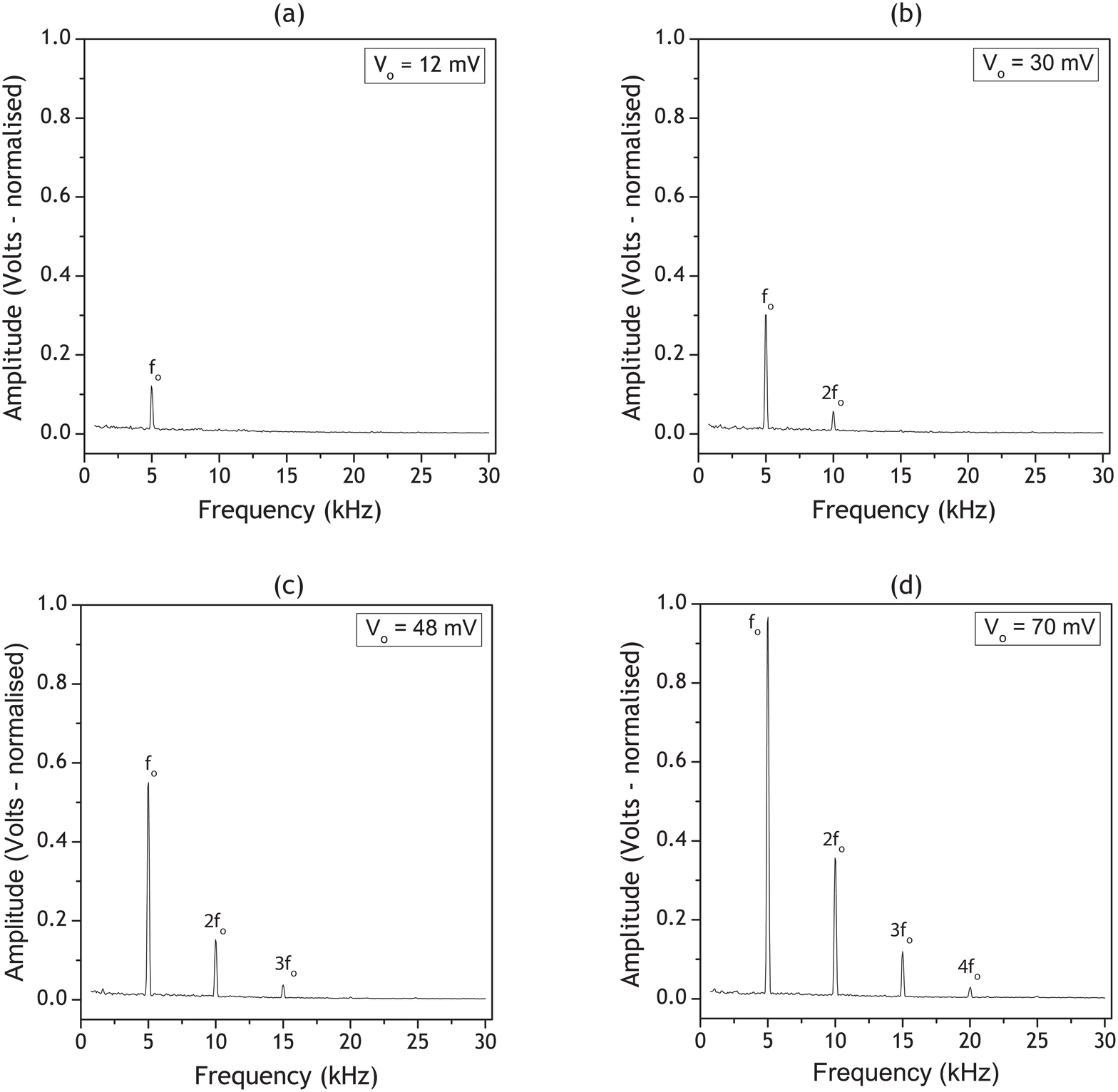}
		\caption{\label{fig:fig4} A series of plots  showing a
			gradual emergence of a nonlinear regime of higher orders as
			the amplitude of the input voltage waveform is increased.
			The amplitude of the input sinusoidal voltage is shown
			on each spectrum. The dc-offset is constant and
			frequency of the input sinusoidal voltage is 5~kHz in all plots. Plot (a) shows
			a linear regime where no new frequency is generated. (b)
			Represents an emergence of the nonlinear regime where the
			second harmonic generation is observed. (c) A third order
			nonlinearity is also pronounced and frequency components up
			to the third harmonics are observed. (d) A fourth order
			nonlinearity is gradually pronounced and a frequency component of 
			fourth harmonics is observed.}
	\end{figure*}
A gradual emergence of a nonlinear regime is
observed by generation of higher order harmonics as the
amplitude of the input sinusoidal voltage $V_{s}$ is
increased. A spectrum analyzer is connected to the
output of the circuit to record the frequency spectrum of the output voltage waveform. The output voltage is also plotted on a digital oscilloscope to record the signal in time domain.  A series of plots shown in Figure.~\ref{fig:fig4}
shows a gradual appearance of higher order harmonics as
the amplitude of the sinusoidal input voltage is increased.
The fundamental frequency of the input sinusoidal voltage waveform
is chosen to be $f_{o}=\omega_{o}/2\pi=5\,kHz$ and the
offset voltage is $V_{dc}$ = 360 mV  in all  the
plots. Figure.~\ref{fig:fig4} (a) corresponds to a linear
regime, where the amplitude of higher harmonics is
negligibly small. A plot in
Figure.~\ref{fig:fig4}(b) shows the emergence of the
nonlinear regime, where the second order nonlinearity
becomes significant,  and  a second harmonics at
frequency 10~$kHz$ is observed.
Another plot in Figure.~\ref{fig:fig4}(c) shows a regime
where the nonlinearity up to the third order is pronounced
as shown by a third harmonic
generation at frequency 15~$kHz$.   A plot in
Figure.~\ref{fig:fig4}(d) shows a regime where the
nonlinearity up to the fourth order is pronounced as
shown by a fourth harmonic generation
at frequency 20~$kHz$.  Therefore, higher orders of nonlinearity can be addressed by increasing
voltage across a $PN$-junction diode.

\subsection{Frequency comb generation}
\label{comb}
\begin{center}
	\begin{figure}
		\includegraphics[scale=0.45]{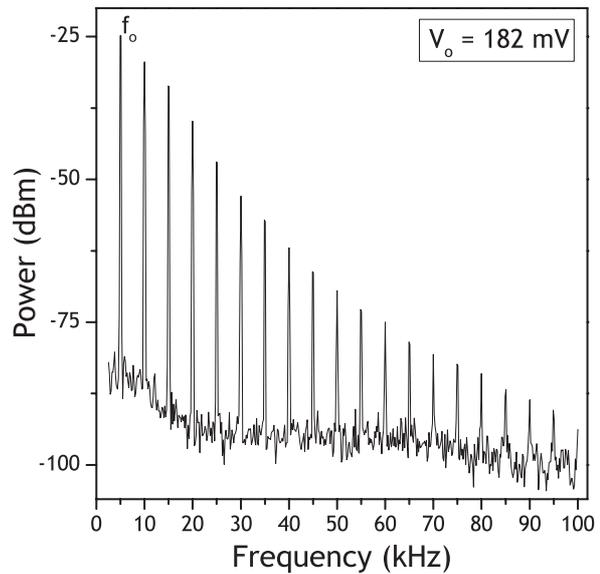}
		\caption{\label{fig:fig5} A frequency spectrum of a
			frequency comb. A highest frequency 100~kHz corresponds to
			the twentieth harmonics and frequency of input voltage
			waveform is $f_{o}=5~kHz$.}
	\end{figure}
\end{center}

A frequency comb is a series of equally spaced discrete
harmonics. In this experiment a frequency comb has been
generated by utilizing the nonlinear response of a $PN$-junction diode. A
sinusoidal input voltage with a nonzero voltage offset is applied
at the input of the circuit as shown in Figure.~\ref{fig:fig2}.
The output voltage is observed in time domain
on a digital oscilloscope. The amplitude of  the input
sinusoidal voltage of frequency 5~$kHz$ is
gradually increased until the output similar to the
output voltage shown in Figure.~\ref{fig:fig3} is
observed.
A frequency spectrum analyzer is connected to the output of
the circuit to measure a frequency spectrum of the output
voltage. For a large amplitude the higher order
nonlinear terms are pronounced and a
comb of frequencies with the highest frequency up to the
twentieth harmonics of frequency 100~$kHz$ is observed,
where the frequency of input sinusoidal voltage is 5~$kHz$.
An electrical power spectrum of the output voltage
is shown in Figure.~\ref{fig:fig5}, where the measured
electrical power is indicated on a logarithmic scale (1~dBm= $10\log_{10}P/1 mW$ and $P$ is measured
in milli-Watt (mW)).  This is essential to keep the
amplitudes corresponding to different frequency components
within the scale margins. Frequency combs are useful in many areas
of science and this demonstration introduces the idea in a
simple system.
\subsection{Sum and difference frequency generation}
\label{sum}
The sum and difference of frequencies of different sinusoidal
voltage waveforms can be produced from the second
order nonlinearity. Two independent sinusoidal voltage
waveforms $V_{s1}$ and $V_{s2}$ with a constant dc voltage
offset are applied at the inputs of the circuit shown in
Figure.~\ref{fig:fig2}. A second sinusoidal voltage
waveform $V_{s2}$ is connected to an inverting terminal of
the first operational amplifier through an additional 10
k$\Omega$ resistance (not shown  in the circuit
diagram). In this case the total input voltage is
$V= V_{dc}+V_{s1}+V_{s2}$ where,
$V_{s1}=V_{o1}\cos(\omega_{1} t)$ and
$V_{s2}=V_{o2}\cos(\omega_{2} t)$. The output voltage is
given by Equation.~\ref{eq:5}. Consider the waveform
amplitudes $V_{o1}$ and $V_{o2}$ are such that only 
first two terms of the Equation.~\ref{eq:5} are significant
\emph{i.e.} only the second order nonlinearity is
significant. Therefore, the total output voltage of  the circuit is
written as
\begin{multline}
\label{eq:9}
V_{t}=R_{F}I_{0}\bigg(\frac{\chi^{(2)}_{v}}{2}(V^{2}_{o1}+V^{2}_{o2}+2
V^{2}_{dc}) +\chi^{(1)}_{v}V_{dc} + \\
(\chi^{(1)}_{v}V_{o1}+ 2\chi^{(2)}_{v}V_{o1}
V_{dc})\cos(\omega_{1}t)+
\\\frac{\chi^{(2)}_{v}}{2}V^{2}_{o1}\cos(2 \omega_{1}t) +
\\(\chi^{(1)}_{v}V_{o2}+2\chi^{(2)}_{v}V_{o2}
V_{dc})\cos(\omega_{2}t)+\\\frac{\chi^{(2)}_{v}}{2}V^{2}_{o2}\cos(2
\omega_{2}t) +
\chi^{(2)}_{v}V_{o1}V_{o2}\cos(\omega_{1}-\omega_{2})t+\\
\chi^{(2)}_{v}V_{o1}V_{o2}\cos(\omega_{1}+\omega_{2})t\bigg)
\end{multline}
\begin{center}
	\begin{figure}
		\includegraphics[scale=0.45]{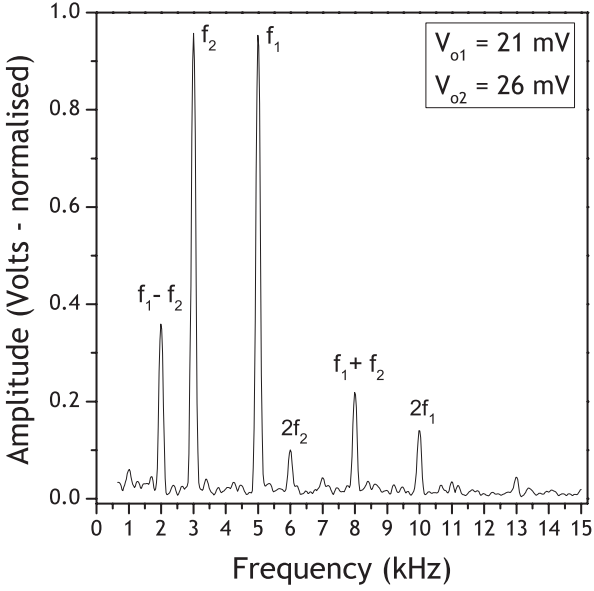}
		\caption{\label{fig:fig6} A frequency spectrum of the output voltage waveform shows the
			sum ($f_{1}+f_{2}$) and difference ($f_{1}-f_{2}$)
			frequency generation.  Second
			harmonics of frequencies $2f_{1}$ and $2f_{2}$ corresponding to each input
			voltage waveforms are also produced.}
	\end{figure}
\end{center}
The output voltage waveform consists of harmonics of
frequencies  equal to the sum ($\omega_{1}+\omega_{2}$)
and the difference ($\omega_{1}-\omega_{2}$) of
frequencies of the input voltage waveforms. In
addition, there are frequency components at $2\omega_{1}$
and $2\omega_{2}$ which correspond to second harmonic
generation corresponding to the individual voltage waveforms
$V_{s1}$ and $V_{s2}$, respectively.
A measured frequency spectrum of the output voltage waveform
$V_{t}$ is shown in Figure.~\ref{fig:fig6} for
$f_{1}=\omega_{1}/2\pi= 5~kHz$ and $f_{2}=\omega_{2}/2\pi
=3~kHz$ where, $V_{o1}= 21~mV$ and $V_{o2} = 26~mV$ are
chosen such that only the second order nonlinearity is
significant. A measured spectrum explicitly shows the sum
of frequencies ($f_{1}+f_{2} = 8~kHz$) and the difference of
frequencies ($f_{1}-f_{2}= 2~kHz$). In addition.  the
measured spectrum contains frequency components of
second harmonic generation at $2f_{1} =10~kHz$ and
$2f_{2}=6~kHz$ corresponding to the individual voltage
waveforms $V_{o1}$ and $V_{o2}$, respectively.

\section{Conclusions}
\label{conc}
In  summary, we have explored the nonlinearity of
voltage current characteristics of a $PN$-junction diode
and experimentally demonstrated various concepts related to
nonlinear physics in particular nonlinear optics. 
Taylor expansion of the voltage current characteristics of a
$PN$-junction diode resembles the nonlinear dependence of
electric polarization of a dielectric medium on the electric
field which allows an analogy between the two
systems.

A gradual emergence of the nonlinear regime of successively
increasing orders has been shown experimentally.  A frequency
comb with the highest frequency up to the twentieth harmonics
has also been produced.  In addition, an experiment to create the sum and
 difference of frequencies of two independent sinusoidal voltage waveforms is
presented. It may also be possible to
observe certain other aspects of nonlinear physics using the $PN$-junction
diode experiment and these aspects will be presented elsewhere.

\end{document}